\documentclass{elsart}

\usepackage{graphicx}
\usepackage{epsfig}

\usepackage{amsmath}
\usepackage{amssymb}

\begin{document}

\begin{frontmatter}



\title{Factorization for hadronic heavy quarkonium 
       production\thanksref{grants}}
\thanks[grants]{Supported in part by the U.S. Department of Energy 
under Grant No. DE-FG02-87ER40371 and Contract No. DE-AC02-98CH10886.}


\author{Jian-Wei Qiu\thanksref{email}}
\thanks[email]{{\it Email address:}\ {\tt jwq@iastate.edu} }

\address{Department of Physics and Astronomy, 
         Iowa State University\\
         Ames, Iowa 50011, U.S.A.\\
         Physics Department, Brookhaven National Laboratory\\
         Upton, New York 11973-5000, U.S.A.}

\begin{abstract}
We briefly review several models of heavy quarkonium 
production in hadronic collisions, and discuss the status of QCD
factorization for these production models. 
\end{abstract}

\end{frontmatter}

\section{Introduction}
\label{intro}

RHIC has produced sufficient evidence that a new state of hot and dense  
matter of quarks and gluons, or the quark-gluon plasma (QGP),
was formed in ultrarelativistic 
heavy ion collisions \cite{rhic-3yrs}.  
Anomalous suppression of heavy quarkonium, J/$\psi$, 
was suggested as a good probe  
for discovering the QGP \cite{MS-jpsi}.  
Because of the heavy mass, production of heavy quark pairs 
takes place at a very short time and 
is unlikely to interfere with the formation of QGP.  
On the other hand, the probability for the heavy quark pair to 
become a bound meson in the QGP could be suppressed 
due to the Debye screening from color charges in the medium.

However, the suppression of heavy quarkonia in a dense medium  
also depends on how the bound state was formed.  A quicker 
formation would give less time for the medium's color charges to
break the coherence of the heavy quark pair, and would require a 
higher color density and/or temperature to produce the suppression. 
It is therefore very important to ask how well we can calibrate 
the suppression in heavy ion collisions to extract useful 
information on medium properties.
In this talk, I first briefly review several models for heavy 
quarkonium production, and then, discuss the status of QCD 
factorization for these production models, and finally, give
a brief summary and outlook.  

\section{Production models}
\label{production}

Heavy quarkonium production has been the subject of a vast 
theoretical literature and of intensive experimental study 
\cite{qwg-review}.
In order to produce a heavy quarkonium in hadronic collisions, 
the energy exchange in the collisions has to be larger
than the invariant mass of the produced quark pair 
($\ge 2m_Q$ with heavy quark mass $m_Q$).  
Since the binding energy of a heavy quarkonium of 
mass $M$ is much less than heavy quark mass, 
$(M^2-4m_Q^2)/4m_Q^2\ll 1$, the 
transition from the pair to a meson is sensitive to 
soft physics.  The quantum interference between the production
of the heavy quark pairs and the transition process 
is  powerly suppressed by the heavy quark mass, and 
the production rate for a heavy quarkonium state, $H$, 
up to corrections in powers of $1/m_{Q}$, 
can  be factorized as, 
\begin{equation}
\sigma_{A+B\rightarrow H+X} 
\approx 
\sum_{n} \int d\Gamma_{Q\bar{Q}}\
\sigma_{A+B\rightarrow Q\bar{Q}[n]+X}(\Gamma_{Q\bar{Q}}, m_Q) \
F_{Q\bar{Q}[n]\rightarrow H}(\Gamma_{Q\bar{Q}}) 
\label{qq-fac}
\end{equation}
with a sum over possible $Q\bar{Q}[n]$ states and an integration 
over available $Q\bar{Q}$ phase space $d\Gamma_{Q\bar{Q}}$.
The nonperturbative transition probability $F$ for a pair of
off-shell heavy quark ($\psi$) and antiquark ($\chi$)
to a quarkonium state $H$ is proportional to the Fourier transform 
of following matrix elements
\begin{equation}
\sum_N
\langle 0|\chi^\dagger(y_1)\, {\mathcal K}_n\, \psi(y_2)|H+N\rangle
\langle H+N| \psi^\dagger(\tilde{y}_2)\, {\mathcal K}'_n\,
             \chi(\tilde{y}_1)|0\rangle\, ,
\label{F-nonlocal}
\end{equation}
where $y_i (\tilde{y}_i)$ are coordinates, and
${\mathcal K}_{n}$ and ${\mathcal K}'_{n}$ are 
local combinations of color and spin matrices for the $Q\bar{Q}$  
state $n$.  A proper insertion of Wilson lines to make
the operators in Eq.~(\ref{F-nonlocal}) gauge invariant is 
implicit \cite{nqs}.  
In Eq.~(\ref{F-nonlocal}),
$\sigma_{A+B\rightarrow Q\bar{Q}[n]+X}$, represents the production 
of a pair of on-shell heavy quarks and is calculable in perturbative 
QCD \cite{heavyquark}.  
The debate on the production mechanism 
has being focused on the transition from the pair to the meson.

\begin{figure}
\begin{minipage}[c]{2.7in}
\begin{center}
  \includegraphics[width=2.4in,height=1.8in]{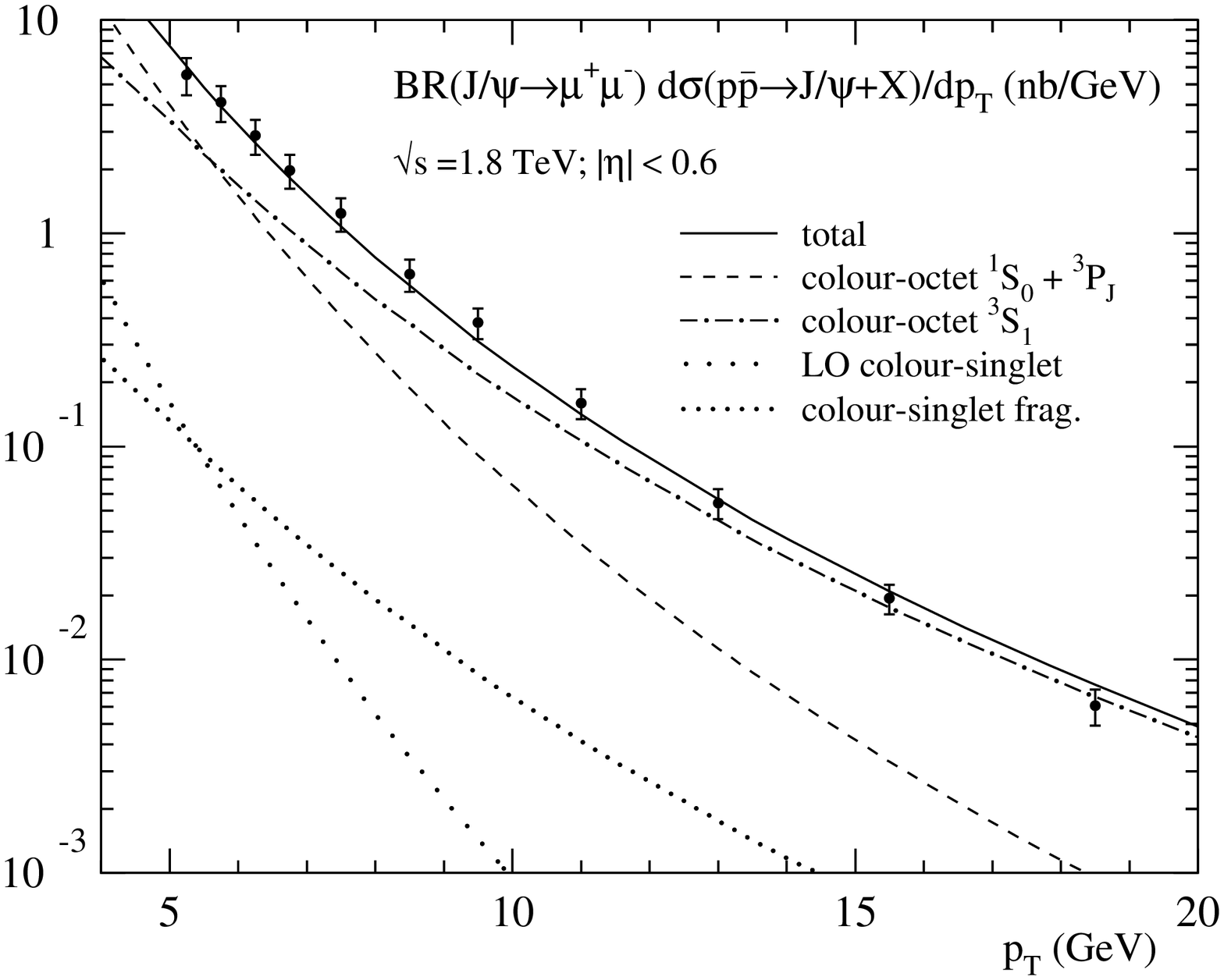}

(a)
\end{center}
\end{minipage}
\hfil
\begin{minipage}[c]{2.7in}
\begin{center}
  \includegraphics[width=1.9in,height=2.6in,angle=270]{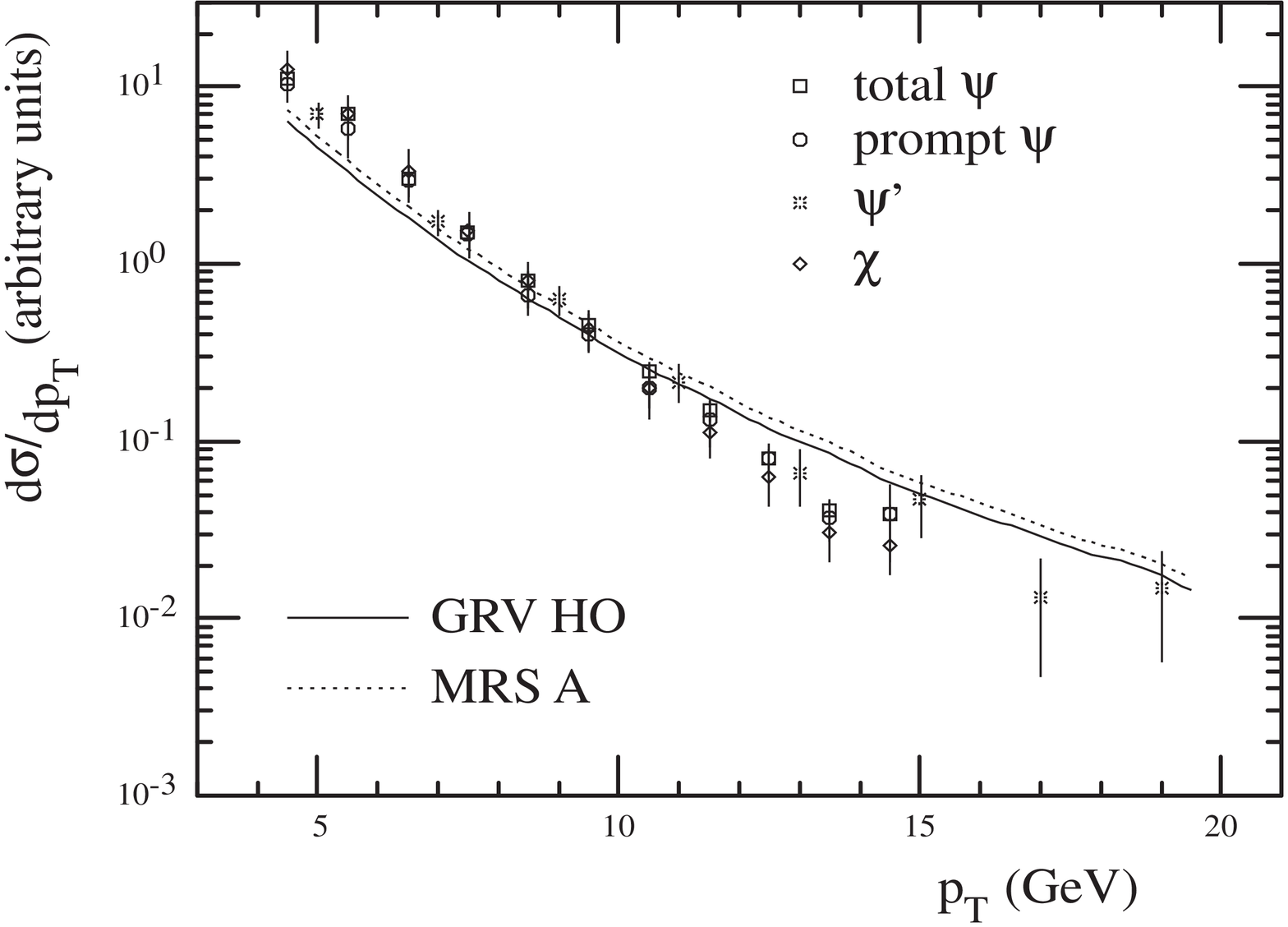}

(b)
\end{center}
\end{minipage}
\caption{Charmonium cross section as a function of $p_T$ along with 
the CDF data points \protect\cite{cdf-jpsi-run1} and 
the theory curves from 
NRQCD model from Ref.~\protect\cite{nrqcd-review} (a), and 
CEM from Ref.~\protect\cite{cem2} (b).}
\label{fig1}
\end{figure}

\noindent{\it Color-singlet model}\ \
The color-singlet model (CSM) assumes that only a color singlet 
heavy quark pair with the right quantum number can become a 
quarkonium of the same quantum number and the transition 
from the pair to a meson is given 
by the quarkonium wave function \cite{csm}.  
$\sigma_{A+B\rightarrow Q\bar{Q}+X}$'s 
dependence on relative momentum of the pair is neglected, and 
$\int d\Gamma_{Q\bar{Q}}\, F_{Q\bar{Q}\to H}$ in Eq.~(\ref{qq-fac})
is equal to the matrix element in Eq.~(\ref{F-nonlocal}) 
evaluated at $y_i (\tilde{y}_i)=0$, which is proportional to the 
square of coordinate-space quarkonium wave function
at the origin, $|R_H(0)|^2$ \cite{qwg-review},
and therefore,
\begin{equation}
\sigma^{\rm CSM}_{A+B\rightarrow H+X} 
\propto
\sigma_{A+B\rightarrow Q\bar{Q}[H]+X}(m_Q)\,
|R_H(0)|^2\, .
\end{equation}
The same wave function appears in both production and decay, and 
the model provides absolutely normalized predictions.  It works well 
for J/$\psi$ production in deep inelastic scattering, photon 
production, and some low energy experiments \cite{qwg-review}, 
but fails to predict the CDF data, 
see the dotted lines in Fig.~\ref{fig1}(a) \cite{nrqcd-review}.

\noindent{\it Color evaporation model}\ \
The color evaporation model (CEM) assumes
that all $Q\bar{Q}$ pairs with invariant mass less than the threshold of 
producing a pair of open-flavor heavy mesons, 
regardless their color, spin, and invariant mass, 
have the same probability to become a quarkonium \cite{cem}.  
That is, the $F_{Q\bar{Q}[n]\to H}$ in Eq.~(\ref{qq-fac}) is 
a constant for a given quarkonium state, $H$, and therefore,
\begin{equation}
\sigma^{\rm CEM}_{A+B\rightarrow H+X} 
\approx 
f_H\, 
\int_{2m_Q}^{2M_Q} dm_{Q\bar{Q}}\
\sigma_{A+B\rightarrow Q\bar{Q}+X}(m_{Q\bar{Q}})
\label{cem-fac}
\end{equation}
with open-flavor heavy meson mass $M_Q$ and a constant $f_H$ \cite{cem}. 
With a proper choice of $m_Q$ and $f_H$, 
the model gives a reasonable description of almost all data 
including the CDF data, as seen in Fig.~\ref{fig1}(b).

\begin{figure}
\begin{minipage}[c]{2.7in}
\begin{center}
  \includegraphics[width=2.6in,height=1.8in]{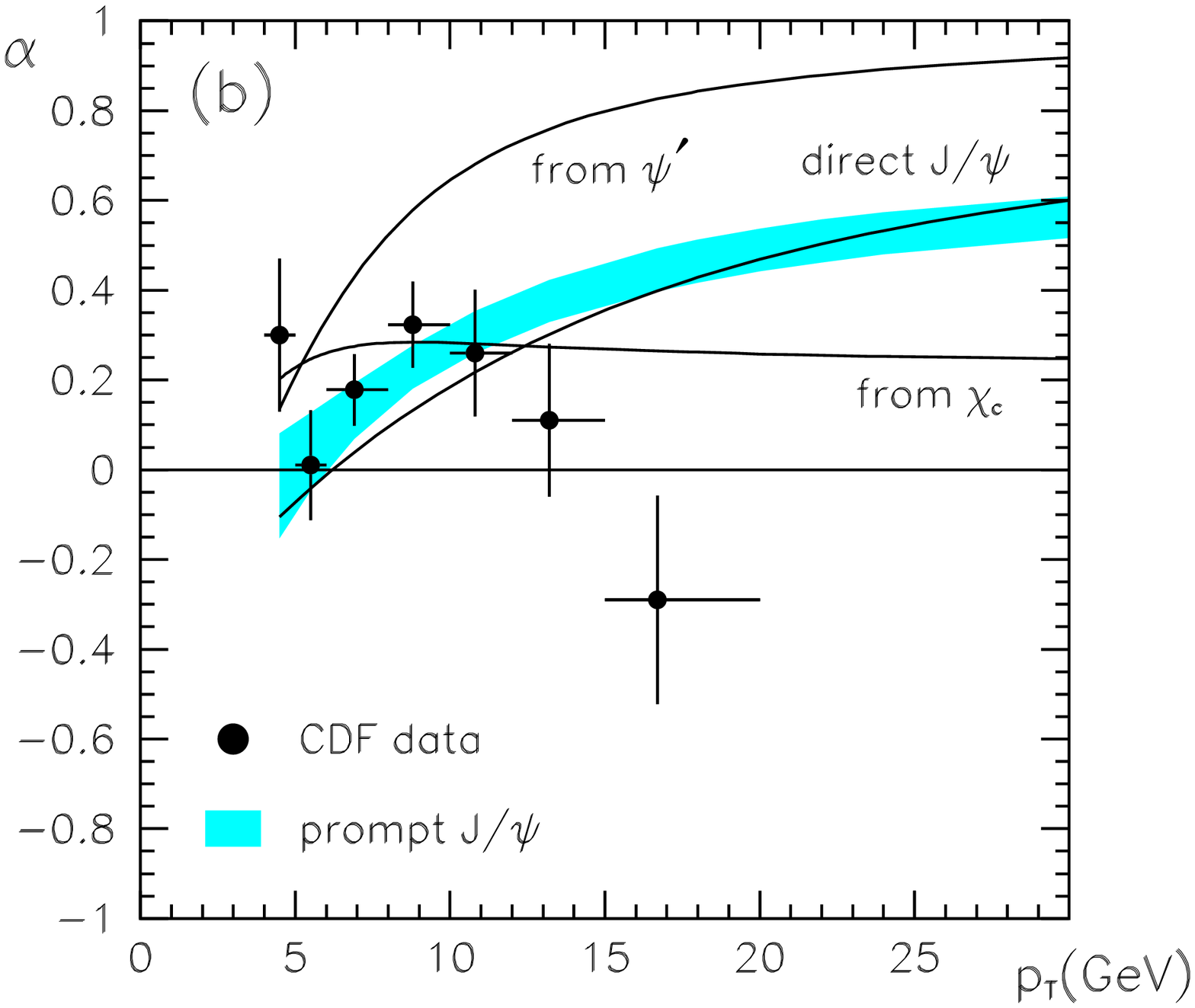}

(a)
\end{center}
\end{minipage}
\hfil
\begin{minipage}[c]{2.7in}
\begin{center}
  \includegraphics[width=2.6in,height=1.8in]{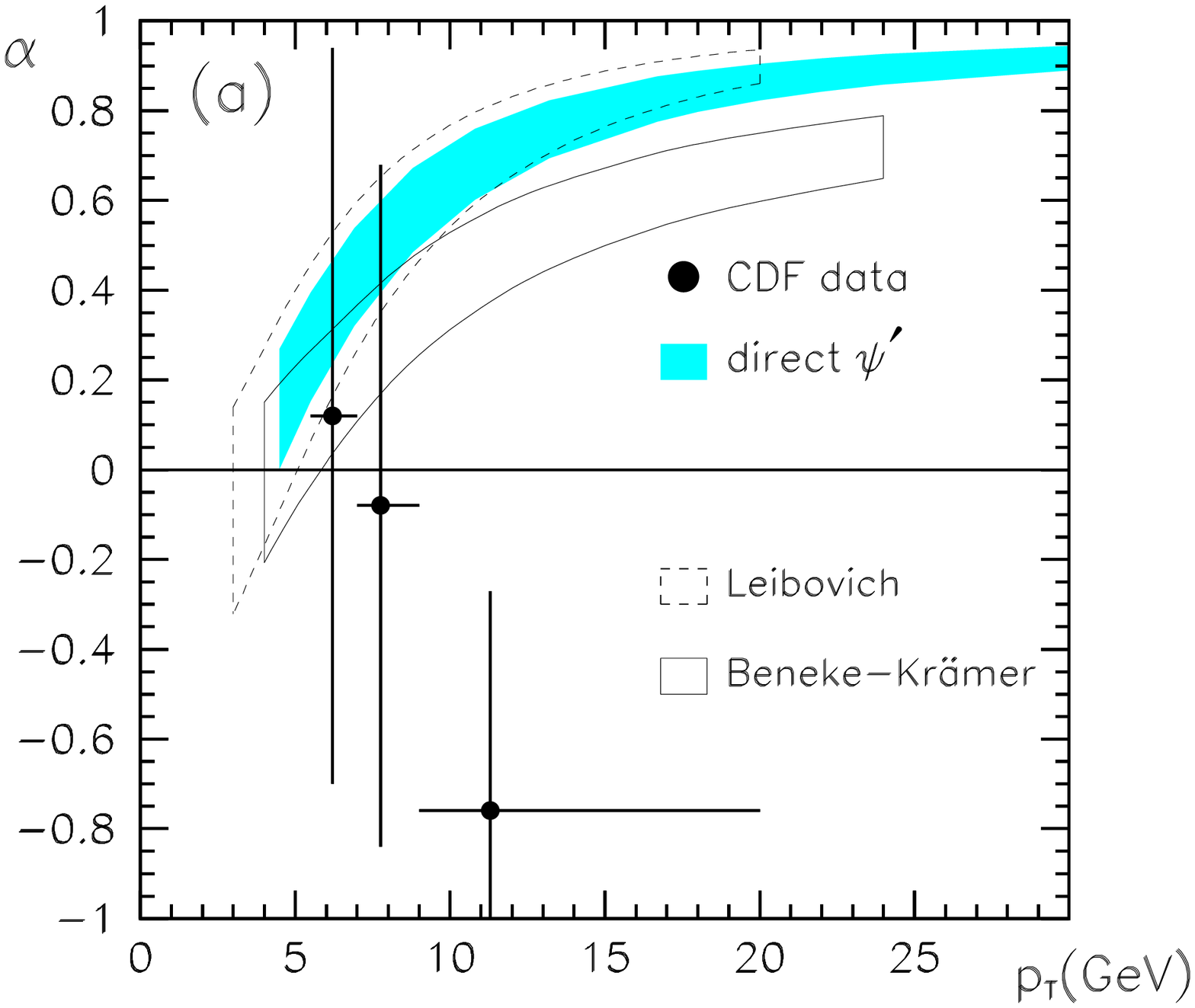}

(b)
\end{center}
\end{minipage}
\caption{From Ref.~\protect\cite{nrqcd-review}, NRQCD 
predictions of charmonium polarizations are compared with  
the CDF data \protect\cite{cdf-pol}.}
\label{fig2}
\end{figure}

\noindent{\it Nonrelativistic QCD model}\ \
The Nonrelativistic QCD (NRQCD) model is based on the fact that
the typical heavy quark rest-frame kinetic energy and binding energy,
$m_Q v^2$, in a heavy quarkonium is much smaller than the heavy quark 
mass.  The model separates the physics at scales of order $m_Q$ and 
higher from the dynamics of the binding by using NRQCD, 
an effective field theory \cite{bbl-nrqcd}.  
It provides a systematic prescription to calculate the physics
at $m_Q$ order-by-order in powers of $\alpha_s$, and
expands the nonperturbative dynamics in terms of local matrix 
elements in power series of heavy quark velocity $v$ 
\cite{nrqcd-review,bbl-nrqcd},
\begin{equation}
\sigma^{\rm NRQCD}_{A+B\to H+X} 
= \sum_n 
\hat\sigma_{A+B\to Q\bar{Q}[n]+X}\
\langle {\mathcal  O}^H_n\rangle\, ,
\label{nrqcd-fac}
\end{equation}
where the ${\mathcal O}^{H}_n$ are NRQCD operators
for the state $H$ \cite{bbl-nrqcd}, 
\begin{equation}
{\mathcal O}^H_n(0)
=
\chi^\dagger(0)\, {\mathcal K}_n\, \psi(0)\, 
\left(a^\dagger_Ha_H\right)\,
\psi^\dagger(0)\, {\mathcal K}'_n\, \chi(0)\, ,
\label{On-def}
\end{equation}
where $a^\dagger_H$ is the creation operator for $H$, 
$\chi$ ($\psi$) are two component Dirac spinors, and 
${\mathcal K}_n$ and ${\mathcal K}'_n$ 
are defined in Eq.~(\ref{F-nonlocal}) and  
can also involve covariant derivatives.  
At higher orders in $v$, the operator ${\mathcal O}^{H}_n$ 
can have additional dependence on field strength as well as more
fermion fields.  The factorization in Eq.~(\ref{nrqcd-fac}) could 
be understood from Eq.~(\ref{qq-fac}) by expanding the
$\sigma_{A+B\rightarrow Q\bar{Q}[n]+X}$ at heavy
quark relative momentum, $q=(p_Q-p_{\bar{Q}})/2=0$. The
moments, $\int d\Gamma_{Q\bar{Q}}\ q^N\, F_{Q\bar{Q}\to H}$,
lead to local matrix elements with high power of $v$. 

The NRQCD model allows every $Q\bar{Q}[n]$ state 
to become a bound quarkonium, 
while the probability is determined by corresponding
nonperturbative matrix elements $\langle {\mathcal  O}^H_n\rangle$.
Its octet contribution
is  the most important one for high $p_T$
quarkonium production at collider energies \cite{nrqcd-review}. 
The NRQCD model has been most successful in 
interpreting data 
\cite{qwg-review,nrqcd-review}, as seen in  Fig.~\ref{fig1}(a).

\noindent{\it Quarkonium polarization and other models}\ \
The key difference between the NRQCD model and the CEM is the 
prediction on quarkonium polarization.  Once the matrix elements
$\langle {\mathcal  O}^H_n\rangle$ are determined, the NRQCD
model can systematically calculate  the polarization
of produced heavy quarkonia. We cannot calculate the 
polarization in CEM because of the uncontrolled radiation 
from the quark pair. 
The polarization of quarkonia at large $p_T$ 
was considered a definite test of the NRQCD model 
\cite{nrqcd-review}.  But, the model has failed the test, 
as seen in Fig.~\ref{fig2}, 
if the CDF data hold \cite{cdf-pol}.  
Several improved and new models have been proposed to 
address the issues of polarization \cite{Lansberg:2006dh}.

\begin{figure}
\begin{minipage}[c]{2.4in}
\begin{center}
  \includegraphics[width=1.7in,height=1.4in]{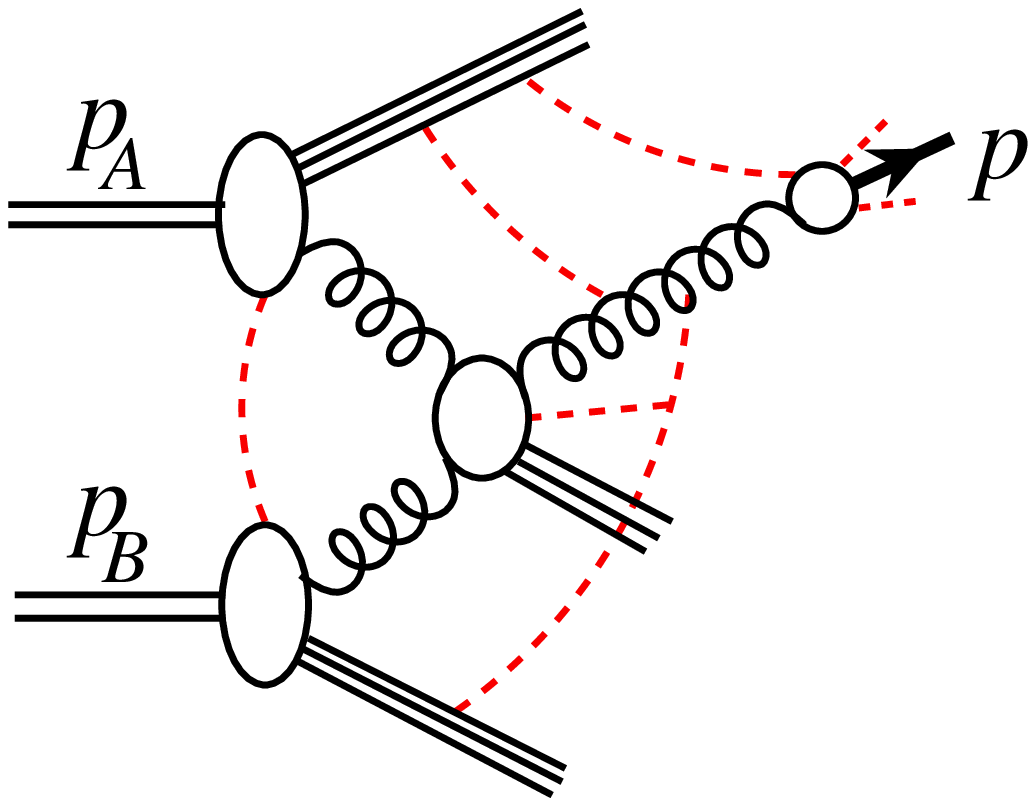}

(a)
\end{center}
\end{minipage}
\hskip 0.2in
\begin{minipage}[c]{2.8in}
\begin{center}
\hskip0.2in
\begin{minipage}[c]{1.0in}
  \includegraphics[width=1.0in,height=0.7in]{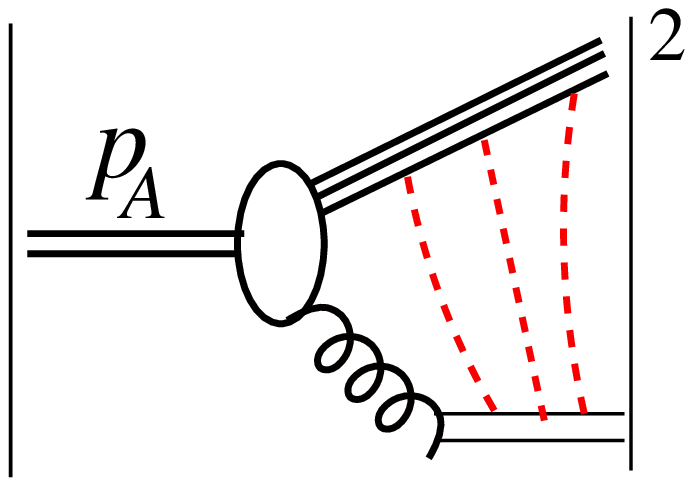}
\end{minipage}
   \hskip 0.1in
   {\large $\otimes $}
   \hskip 0.1in
\begin{minipage}[c]{1.0in}
  \includegraphics[width=1.0in,height=0.7in]{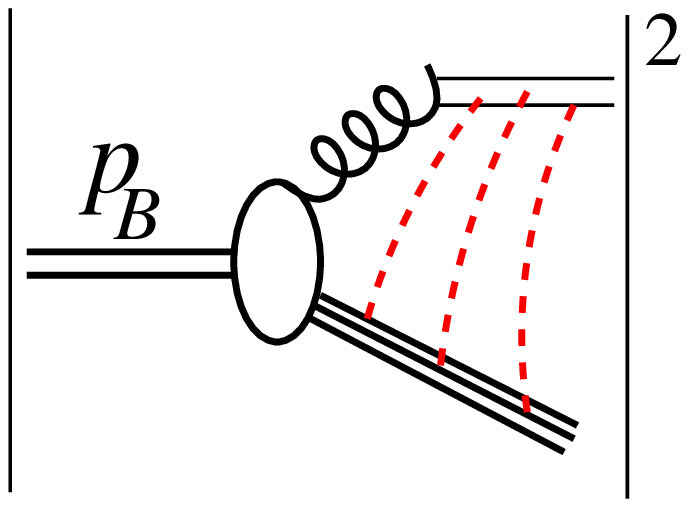}
\end{minipage}

\vskip 0.1in

   {\large $\otimes $}
   \hskip 0.1in
\begin{minipage}[c]{0.8in}
  \includegraphics[width=0.8in,height=0.6in]{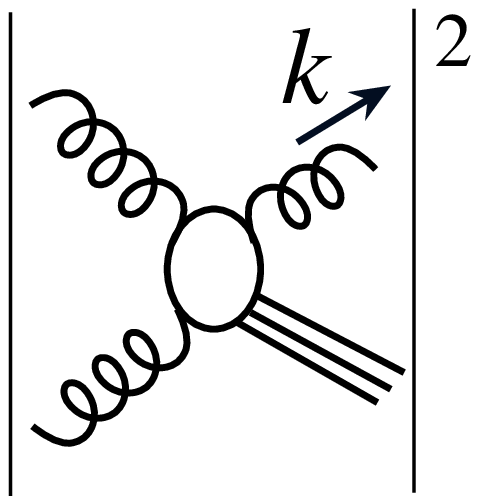}
\end{minipage}
   \hskip 0.1in
   {\large $\otimes $}
   \hskip 0.1in
\begin{minipage}[c]{0.8in}
  \includegraphics[width=0.8in,height=0.6in]{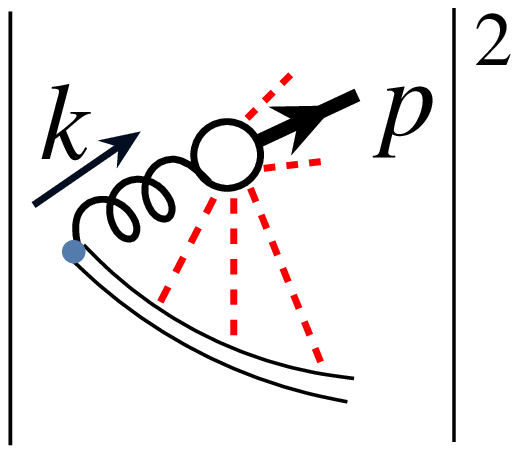}
\end{minipage}

\hskip 0.2in (b)  
\end{center}
\end{minipage}
\caption{Sketch for process $A+B\to H(p)+X$: 
(a) sample scattering amplitude with 
    possible factorization breaking soft interactions 
    indicated by dashed lines, and 
(b) factorization with Wilson lines 
    indicated by thin double lines.}
\label{fig3}
\end{figure}

\section{Factorization}
\label{factorization}

Heavy quarkonium production in hadronic collisions involves both 
perturbative and nonperturbative scales.  Nonperturbative physics 
appears not only in the transition from the heavy quark pair to 
a bound state but also in incoming hadron wave functions.
A typical scattering amplitude for quarkonium production, 
as sketched in Fig.~\ref{fig3}(a), can have soft and 
nonperturbative interactions between incoming hadrons 
as well as between the spectators and the formation process.
These soft interactions may introduce 
process dependence to the nonperturbative matrix elements, and
consequently, spoil the predictive powers of
Eqs.~(\ref{cem-fac}) and (\ref{nrqcd-fac}). 
 
A proof of the factorization needs to: 1) show that the square of the 
scattering amplitude in Fig.~\ref{fig3}(a), after summing over 
all amplitudes with the same initial and final states, 
can be expressed as a convolution of the probabilities, 
as sketched in Fig.~\ref{fig3}(b); each probability represents 
a square of sub-amplitudes and is evaluated at its own
momentum scale(s); 2) show that the piece evaluated at 
perturbative scale(s) is infrared safe and those evaluated 
at nonperturbative scales are universal.  

As argued in Ref.~\cite{heavyquark}, cross sections for producing
on-shell heavy quark pairs can be computed in terms of QCD 
factorization.  Therefore, the right-hand-side of the 
CEM formalism in Eq.~(\ref{cem-fac}) is perturbatively 
calculable for quarkonium production at a large $p_T$, 
and the $f_H$ should be universal {\it within the model}\ because
 soft interactions in Fig.~\ref{fig3}(a)
are suppressed by powers of $1/p_T$.  However, the factorization
statement does not provide justification  that the transition from 
a heavy quark pair to a quarkonium 
state is independent of the pair's invariant mass 
$m_{Q\bar{Q}}$, spin, and other quantum numbers.  
When $p_T \ll m_Q$, the universality of $f_H$ may not  
be valid because of the spectator interactions.

Fully convincing arguments
have not yet been given for NRQCD factorization formalism in
Eq.~(\ref{nrqcd-fac}) \cite{nrqcd-review,bbl-nrqcd}.  
Since the spectator interaction between the beam jet 
and the jet of heavy quark pair should be suppressed by powers of 
$1/p_T$, one might expect the NRQCD factorization formalism to work 
at large $p_T$.  
When $p_T\gg m_Q$, heavy quarkonium production is similar to the 
single light hadron production, and is dominated by parton 
fragmentation.  The cross section is 
proportional to the universal parton-to-hadron fragmentation 
functions \cite{nqs},
\begin{equation}
\sigma_{A+B\to H+X}(p_T) = 
\sum_i\; \hat\sigma_{A+B\to i+X}(p_T/z,\mu) \otimes
D_{H/i}(z,m_Q,\mu) + {\mathcal O}(m_H^2/p_T^2)\, .
\label{cofact}
\end{equation}
Here, $\otimes$ represents a convolution in the momentum fraction $z$.
The cross section $\hat\sigma_{A+B\to i+X}$ includes all 
information on the incoming state, including convolutions with
parton distributions of hadrons $A$ and $B$ 
at factorization scale $\mu$, as sketched in 
Fig.~\ref{fig3}(b).  As a necessary condition for
NRQCD factorization in Eq.~(\ref{nrqcd-fac}), 
the following factorization relation, 
\begin{equation}
D_{H/i}(z,m_Q,\mu) = \sum_n\; d_{i\to Q\bar{Q}[n]}(z,\mu,m_Q) \, 
\langle {\mathcal O}^H_n\rangle\, ,
\label{frag-fac}
\end{equation}
is required to be valid to all orders in $\alpha_s$ and 
all powers in $v$-expansion
for all parton-to-quarkonium fragmentation functions \cite{nqs}.
In Eq.~(\ref{frag-fac}), $d_{i\to Q\bar{Q}[n]}$ 
describes the evolution of an off-shell parton into a heavy quark 
pair in state $[n]$, including logarithms of $\mu/m_Q$, and 
should be infrared safe \cite{nqs}.
   
\begin{figure}
\begin{minipage}[c]{1.7in}
\begin{center}
  \includegraphics[width=1.5in,height=1.2in]{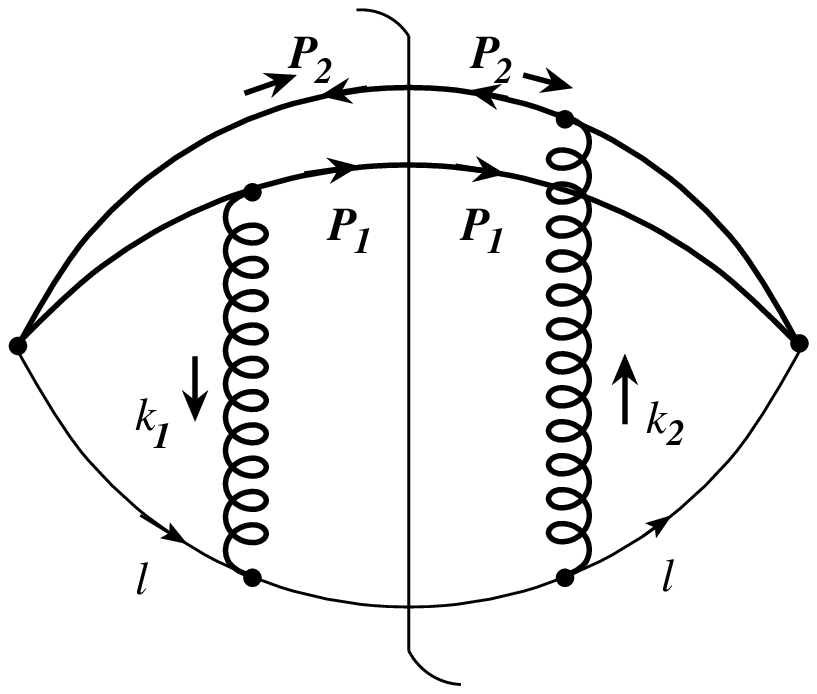}

(I)
\end{center}
\end{minipage}
  \hfil
\begin{minipage}[c]{1.7in}
\begin{center}
  \includegraphics[width=1.5in,height=1.2in]{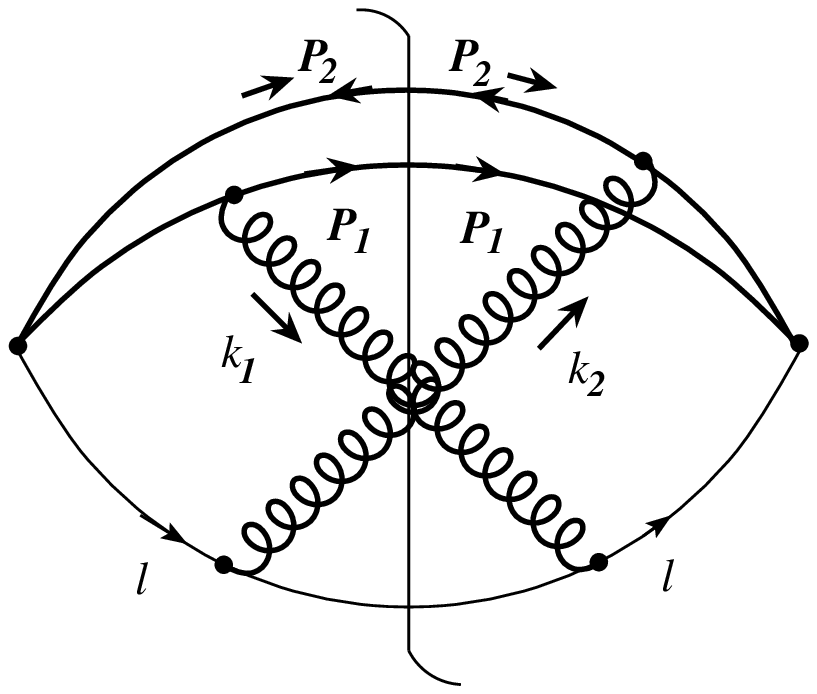}

(II)
\end{center}
\end{minipage}
  \hfil
\begin{minipage}[c]{1.7in}
\begin{center}
  \includegraphics[width=1.5in,height=1.2in]{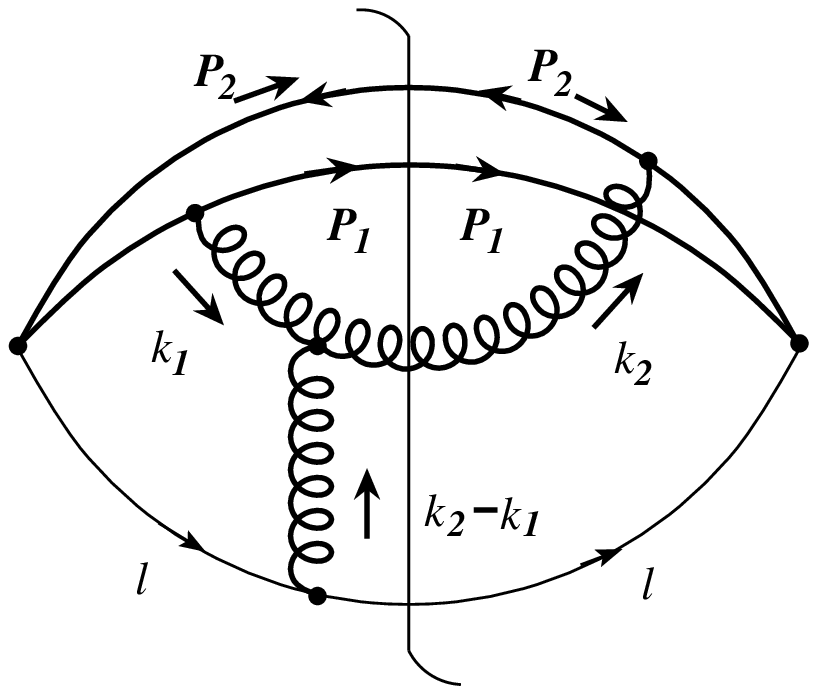}

(III)
\end{center}
\end{minipage}
\caption{Representative NNLO contributions to $g\to Q\bar{Q}$ 
fragmentation in eikonal approximation, see Ref.~\protect\cite{nqs}
for the details.}
\label{fig4}
\end{figure}

The factorization relation in Eq.~(\ref{frag-fac}) 
was tested up to next-to-next-to-leading order (NNLO) in
$\alpha_s$ at $v^2$ order in Ref.~\cite{nqs}, as well as 
at finite $v$ in Ref.~\cite{nqs2}.  
Consider representative NNLO contributions to 
the fragmentation process of transforming a color octet heavy
quark pair to a singlet, as sketched in Fig.~\ref{fig4}.
The individual classes of diagrams in Fig.~\ref{fig4}(I) and (II), 
for which two gluons are exchanged between the
quarks and the Wilson line, satisfy the infrared
cancellation conjecture of Ref.~\cite{bbl-nrqcd}, 
by summing over the possible cuts and connections to quark and
antiquark lines, as do diagrams that have three gluon-eikonal
vertices on the quark pair and one on the  Wilson line \cite{nqs}.  
For Fig.~\ref{fig4}(III) type of diagrams, however, 
with a three-gluon interaction, this cancellation
fails.  By summing over all contributions to second order 
in the relative momentum $q$, it was found that  
the fragmentation function has a noncanceling real pole \cite{nqs}
\begin{equation}
{\mathcal I}^{(8\to 1)}_2(v) 
= - \alpha_s^2\, \frac{1}{3\varepsilon}\, v^2 \, .
\label{v2}
\end{equation}
From Fig.~\ref{fig4}(III), 
this infrared divergence
is not topologically factorizable and 
can not be absorbed into the matrix elements 
$\langle {\mathcal  O}^H_n\rangle$.   
As demonstrated in Ref.~\cite{nqs}, 
as a necessary condition for restoring the NRQCD factorization
at NNLO, the conventional ${\mathcal O}(v^2)$
octet NRQCD production matrix elements
$\langle {\mathcal  O}^H_n\rangle$
must be modified by incorporating Wilson lines 
that make them manifestly gauge invariant,
so that the infrared divergence in Eq.~(\ref{v2}) can be
absorbed into the gauge-completed matrix elements.

For quarkonium production, all heavy quark pairs with invariant mass 
less than a pair of open-flavor heavy mesons could become
a bound quarkonium.  Therefore, the effective velocity of heavy
quark pair in the production could be much larger than that in decay.
For charmonium production, 
charm quark velocity, $v_c\sim |\vec{q}_c|/m_c \leq
\sqrt{(4M_D^2-4m_c^2)/(4m_c^2)} \sim 0.88$, is not small, and 
therefore, the velocity expansion for charmonium production 
may not be a good approximation, unless one can identify and resum
large contributions to all order in $v$ or have a factorized 
formalism at finite $v$.  It was found in Ref.~\cite{nqs2} that
with the gauge-completed matrix elements, 
infrared singularities in the fragmentation function for a color
octet pair to a singlet at NNLO are consistent with NRQCD 
factorization to all orders in $v$ or to a finite $v$,
\begin{equation}
{\mathcal I}^{(8\to 1)}(v)
= \frac{\alpha_s^2}{4\varepsilon} 
\left[ 1-\frac{1}{2f(v)} 
         \ln\left[\frac{1+f(v)}{1-f(v)}\right]
\right] \, ,
\label{allorder}
\end{equation}
where $f(v)=2v/(1+v^2)$.
The result in finite $v$ is remarkably compact and intriguing,
and should encourage further work on the factorization 
theorem.

\section{Summary and outlook}
\label{summary}

Heavy quarkonium has two intrinsic scales and could be a good probe
for the properties of the QGP or other dense medium.  However, 
after more than 30 years since the discovery of J/$\psi$, we still
have not been able to fully understand the production mechanism of 
heavy quarkonium in high energy collisions, in particular, the 
transition from the heavy quark pair to a bound quarkonium.  
Although there are good reasons for each production model, 
none of the factorized production formalism, including that of the 
NRQCD model, has been proved theoretically. Further work on the
factorization theorem is critical.

The transition from the pair to a bound meson is sensitive
to the QCD confinement, and it is the dynamics of the transition that 
interacts and probes the properties of dense medium in heavy ion
collisions.    
Nuclear matter could be an effective filter 
to distinguish the production mechanism  \cite{qvz-jpsi}.  
Detailed study of nuclear dependence of heavy quarkonium 
production in hadron-nucleus collisions 
should provide  invaluable information on the 
formation of heavy quarkonia in hadronic collisions.  
How good a probe the heavy quarkonium production is 
completely depends on how well we understand and are able to 
calibrate the production.





\begin{thebibliography}{00}




\bibitem{rhic-3yrs}
  I.~Arsene {\it et al.}  [BRAHMS Collaboration],
  Nucl.\ Phys.\ A {\bf 757}, 1 (2005);
  B.~B.~Back {\it et al.},
  Nucl.\ Phys.\ A {\bf 757}, 28 (2005);
  J.~Adams {\it et al.}  [STAR Collaboration],
  Nucl.\ Phys.\ A {\bf 757}, 102 (2005);
  K.~Adcox {\it et al.}  [PHENIX Collaboration],
  Nucl.\ Phys.\ A {\bf 757}, 184 (2005).

\bibitem{MS-jpsi} 
  T.~Matsui and H.~Satz,
  Phys.\ Lett.\ B {\bf 178}, 416 (1986).

\bibitem{qwg-review} 
  N.~Brambilla {\it et al.} [Quarkonium Working Group], 
  arXiv:hep-ph/0412158.

\bibitem{nqs} 
  G.~C.~Nayak, J.~W.~Qiu and G.~Sterman,
  Phys.\ Lett.\ B {\bf 613}, 45 (2005);
  Phys.\ Rev.\ D {\bf 72}, 114012 (2005).

\bibitem{heavyquark}
  J.~C.~Collins, D.~E.~Soper and G.~Sterman,
  Nucl.\ Phys.\ B {\bf 263}, 37 (1986).

\bibitem{csm}
  C.~H.~Chang,
  Nucl.\ Phys.\ B {\bf 172}, 425 (1980);
  R.~Baier and R.~Ruckl,
  Phys.\ Lett.\ B {\bf 102}, 364 (1981);
  E.~L.~Berger and D.~L.~Jones,
  Phys.\ Rev.\ D {\bf 23}, 1521 (1981).

\bibitem{nrqcd-review}
  E.~Braaten, S.~Fleming and T.~C.~Yuan,
  Ann.\ Rev.\ Nucl.\ Part.\ Sci.\  {\bf 46}, 197 (1996);
  M.~Kramer,
  Prog.\ Part.\ Nucl.\ Phys.\  {\bf 47}, 141 (2001);
  G.~T.~Bodwin,
  Int.\ J.\ Mod.\ Phys.\ A {\bf 21}, 785 (2006).

\bibitem{cdf-jpsi-run1}
F.~Abe {\it et al.}  [CDF Collaboration],                                
{\it Phys.\ Rev.\ Lett.\ }  {\bf 79}, 572 (1997). 

\bibitem{cem}
  H.~Fritzsch,
   ``Producing Heavy Quark Flavors In Hadronic Collisions: A Test Of Quantum
  Phys.\ Lett.\ B {\bf 67}, 217 (1977);
  F.~Halzen,
  Phys.\ Lett.\ B {\bf 69}, 105 (1977);
  M.~Gluck, J.~F.~Owens and E.~Reya,
  Phys.\ Rev.\ D {\bf 17}, 2324 (1978).

\bibitem{cem2}
  J.~F.~Amundson, O.~J.~P.~Eboli, E.~M.~Gregores and F.~Halzen,
  Phys.\ Lett.\ B {\bf 390}, 323 (1997).

\bibitem{bbl-nrqcd}
  G.~T.~Bodwin, E.~Braaten and G.~P.~Lepage,
  Phys.\ Rev.\ D {\bf 51}, 1125 (1995)
  [Erratum-ibid.\ D {\bf 55}, 5853 (1997)].

\bibitem{cdf-pol}
  A.~A.~Affolder {\it et al.}  [CDF Collaboration],
  Phys.\ Rev.\ Lett.\  {\bf 85}, 2886 (2000);
  CDF Collaboration, Notes 8212 and 8424 (06-06-22).


\bibitem{Lansberg:2006dh}
  J.~P.~Lansberg,
  Int.\ J.\ Mod.\ Phys.\ A {\bf 21}, 3857 (2006)
  [arXiv:hep-ph/0602091].

\bibitem{nqs2}
  G.~C.~Nayak, J.~W.~Qiu and G.~Sterman,
  Phys.\ Rev.\ D {\bf 74}, 074007 (2006).

\bibitem{qvz-jpsi}
  for example, see\ 
  J.~W.~Qiu, J.~P.~Vary and X.~f.~Zhang,
  Phys.\ Rev.\ Lett.\  {\bf 88}, 232301 (2002);
  D.~Kharzeev and K.~Tuchin,
  Nucl.\ Phys.\ A {\bf 770}, 40 (2006);
  and references therein.

\end{thebibliography}
\end{document}